\documentstyle[11pt,aaspp4]{article}
\def\Mdot{\dot M}

\begin{document}

\title{Does the Mass Accretion Rate Depend On the Radius of
the Accreting Star?}

\author{Abraham Loeb, Ramesh Narayan, and John C. Raymond}
\affil{Harvard-Smithsonian Center for Astrophysics, 60 Garden St.,
Cambridge, MA 02138} \affil{\tt e-mail: aloeb@cfa.harvard.edu,
rnarayan@cfa.harvard.edu, jraymond@cfa.harvard.edu}

\begin{abstract}

In some circumstances, the mass accretion rate $\Mdot_\star$ onto a
compact star may depend not only on external boundary conditions, but
also on the radius $R_\star$ of the star.  Writing the dependence as
$\Mdot_\star\propto R_\star^p$, we estimate $p$ for transient binary
systems in a quiescent state.  We use the observed luminosities $L$ of
accreting neutron stars ($R_\star\sim 10^6~{\rm cm}$) in soft X-ray
transients and white dwarfs ($R_\star\sim 10^9~{\rm cm}$) in similar
cataclysmic variables, and estimate $\Mdot_\star$ in each system
through the relation $L\approx GM_\star\Mdot_\star/R_\star$, where
$M_\star$ is the mass of the star.  From the available data we infer
that $p\sim 0.9\pm 0.5$.  This radial dependence is consistent with
radiatively inefficient accretion flows that either are convective or
lose mass via a wind.

\end{abstract}

\section{Introduction}

For steady accretion onto a compact object, it is often assumed that
the mass accretion rate onto the central star $\Mdot_\star$ is
determined only by boundary conditions at some large radius where the
gas is initially introduced into the accretion flow.  This assumption
is valid for a number of idealized accretion solutions: (i) spherical
accretion (Bondi 1952), (ii) steady accretion via a thin disk (Shakura
\& Sunyaev 1973; Novikov \& Thorne 1973), and (iii) steady accretion
via an advection-dominated accretion flow (ADAF; Ichimaru 1977;
Narayan \& Yi 1994; 1995a,b; Abramowicz et al. 1995; Chen et
al. 1995).

Recently, it has become clear that under some circumstances $\dot
M_\star$ may depend not only on outer boundary conditions but also on
the radius of the central star $R_\star$.  A power-law dependence has
been proposed:
\begin{equation}
\Mdot_\star \propto R_\star^{p}.
\end{equation}
Gruzinov (1998), for instance, argued that spherical accretion in the
presence of strong conduction behaves according to equation (1), with
$p$ of order unity.

In the case of a thin accretion disk, it has been known for a number
of years that large changes in the opacity of the accreting gas as a
result of hydrogen ionization lead to a thermal instability (see
Cannizzo 1993 for a review).  This instability causes the well-known
transient behavior of accreting white dwarfs (WDs) in cataclysmic
variables (CVs), and accreting neutron stars (NSs) and black holes in
soft X-ray transients (SXTs).  If accretion in the quiescent state of
transient binaries occurs via a pure thin disk (which is in some
doubt, see below), then $\Mdot_\star$ is expected to scale according
to equation (1) with a large value of $p\sim2.7$ (Ludwig,
Meyer-Hofmeister \& Ritter 1994; Hameury et al. 1998; Menou et
al. 1999a).

In the case of advection-dominated accretion, the radius of the star
can again be important in some circumstances.  For instance, if there
are strong outflows, as seems likely (Narayan \& Yi 1994, 1995a), the
mass accretion rate is expected to behave as in equation (1).
Blandford \& Begelman (1999) suggested that ADAFs with outflows, which
they named ADIOS, may have values of $p$ in the range $0-1$.
Igumenshchev \& Abramowicz (1999, 2000) found ADIOS-like flows in
numerical simulations when they used large values of the viscosity
parameter $\alpha>0.3$.  Their simulations suggest that
$p\sim0.3-0.5$; this estimate is, however, very approximate since the
results from the simulations are not consistent with a power-law
behavior.

Narayan \& Yi (1994, 1995a) argued that advective flows should, in
addition to having outflows, also be convectively unstable.
Igumenshchev \& Abramowicz (1999, 2000) found well-developed
convection in numerical simulations when they used small values of
$\alpha<0.1$.  These convective flows correspond to a new form of
quasi-spherical accretion (Stone, Pringle \& Begelman 1999), which is
called a convection-dominated accretion flow or CDAF.  Both the
numerical simulations and analytical work (Narayan, Igumenshchev \&
Abramowicz 2000; Quataert \& Gruzinov 2000; Igumenshchev, Abramowicz
\& Narayan 2000) show that CDAFs follow equation (1) with $p\sim1$.

In this {\it Letter} we are concerned with transient binaries such as
CVs and SXTs.  Narayan, McClintock \& Yi (1996) and Lasota, Narayan \&
Yi (1996) argued that quiescent SXTs cannot have pure thin disks; a
similar argument had been made earlier by Meyer \& Meyer-Hofmeister
(1994) for CVs.  Narayan et al. (1996) suggested that SXTs, and by
extension CVs, have composite accretion flows in which a thin disk is
present at large radii,  beyond $\sim 10^9-10^{10}$ cm, and an
ADAF is present at smaller radii.  In view of the recent work
described above, it is possible that, rather than an ADAF, the inner
flow might consist of an ADIOS or a CDAF.  We thus have four different
possibilities for the accretion flow in quiescent CVs and SXTs, each
with a different prediction for $p$; (i) pure thin disk ($p=2.7$),
(ii) ADAF ($p=0$), (iii) ADIOS ($p\sim0.3-0.5$), 
and (iv) CDAF ($p\sim1$).

Theoretical analysis is presently incapable of discriminating among
these four possibilities.  Here we use a novel empirical 
method to infer the value of $p$.

We consider CVs and SXTs with similar orbital periods (in the range
$\sim2-20$ hours), which ensures that the outer boundary conditions of
their accretion flows are similar.  The mass transfer rate from the
companion star in a semi-detached binary system depends primarily on the
binary period, and is fairly insensitive to the mass of the compact star
(Menou et al. 1999a).  This is especially true for systems with orbital
periods of a fraction of a day or shorter, in which mass transfer is driven
primarily by gravitational radiation losses.  Thus, for the CVs and SXTs
considered here, the mass supply rate is likely to be similar.

On the other hand, the radius of the accreting star in the systems we
consider spans a wide range, from $\sim 10^9~{\rm cm}$ for WDs in CVs
to $\sim 10^6~{\rm cm}$ for NSs in SXTs.  For accretion onto a star
with a hard surface (as distinct from a black hole, see Narayan,
Garcia \& McClintock 1997), the accretion luminosity $L$ has a simple
dependence on $R_\star$ and $\Mdot_\star$:
\begin{equation}
L\approx  {GM_\star\Mdot_\star \over R_{\star}}~,
\label{eq:1}
\end{equation}
where $M_\star$ is the mass of the accreting star.
Therefore, given the observed $L$, and estimates of $M_\star$ and
$R_\star$, we may infer the value of $\Mdot_\star$.  Since we select
systems with similar outer boundary conditions, but with a wide range
of $R_\star$, we should be able to estimate the dependence of
$\Mdot_\star$ on $R_\star$, and thus the value of $p$.  Note that
correction factors of order unity are unimportant in this analysis,
because the value of the gravitational potential $GM_\star/R_\star
c^2$, or equivalently $L/\Mdot_\star c^2$, differs by three orders of
magnitude between NSs and WDs.

Before proceeding, we note that Medvedev \& Narayan (2000) recently
obtained a new hot quasi-spherical accretion solution which applies
under some circumstances to a rotating star with a surface.  The
luminosity of this solution does not scale according to equation (1),
but depends primarily on the radius and the spin of the star.  The
estimate of $p$ derived in this {\it Letter} would not be valid if
accretion in the CVs or NS SXTs we consider is described by this
solution.

In \S 2 we present the available data on the luminosities of WDs (\S
2.1) and NSs (\S 2.2) with similar binary periods and companions. In
\S 3, we use the data to infer the best-fit value of $p$. Finally, we
summarize our conclusions and discuss their implications in \S 4.

\section{Luminosity Data}

\subsection{White Dwarfs}

CVs have been extensively studied at X-ray, UV and optical
wavelengths.  In recent years, UV spectroscopy has provided WD
temperatures for a number of systems.  The sample of NS SXTs discussed
in \S2.2 has orbital periods in the range $2-20$ hours, with a
geometric mean of about 6 hours.  CVs typically have somewhat shorter
orbital periods.  In order to make a fair comparison, we focus our
attention on non-magnetic CVs with periods greater than 3 hours.  Most
of the long period systems are high accretion rate Nova-like variables
whose UV spectra are dominated by their accretion disks and whose WD
temperatures are not known.  There are six CVs with low $\Mdot$ which
are suitable for the comparison (Table 1).  For the basic parameters,
we take periods ($P$), WD masses ($M_{wd}$), 
V-band luminosities ($L_v$) 
and distances ($d$) from Ritter \&
Kolb (1999), WD temperatures ($T_{wd}$)  from Sion (1999) 
and X-ray luminosities ($L_x$)
from the ROSAT All Sky Survey tabulation of Verbunt et al. (1997).  WD
radii ($R_{wd}$)
are derived from the masses and the mass-radius relation (Hamada
\& Salpeter 1961).  The six selected systems have orbital 
periods in
the range 3.3 to 6.6 hours, with a geometric mean of about 5 hours,
not very different from the mean period of the NS SXTs (6 hours).

We have substituted more recent parameter determinations in several cases.
The distances for SS Cyg and U Gem are those based on HST FES parallaxes
(Harrison et al 1999), which are more reliable than other distance
estimates.  The X-ray luminosities of SS Cyg and U Gem are derived by
scaling the GINGA spectrum of Yoshida, Inoue \& Osaki (1992) and the ASCA
spectrum of Szkody et al. (1996) to the Harrison et al. (1999) distances,
because luminosities based on ROSAT are less reliable for hard sources.
The WD temperature, mass and radius for U Gem are taken from Long (2000),
who finds that a combination of 85\% of the surface of the star at 30,000 K
and 15\% at 50,000 K fits the UV spectrum.  This results in the WD
luminosity quoted in Table 1.  The limit on the WD temperature 
for IP Peg is from Froning et al. (1999), but uncertainties in the
luminosity are relatively high for this edge-on system.  The distance limit
for TZ Per is from Ringwald (1995).

DW UMa is a special case.  Knigge et al. (2000) present a detailed
analysis of HST spectra.  By chance, the observations occurred during
a remarkable low state for this SW Sex star, when it was 3 magnitudes
fainter than its normal brightness.  Knigge et al. (2000) were able to
derive the temperature, mass and radius of the WD, but no X-ray
observations in the low state are available. Moreover, WDs in CVs are
heated during high $\Mdot$ states and cool gradually after $\Mdot$
drops, with characteristic times of months (see the review by Sion
1999).  Therefore, it is probable that DW UMa was cooling from its
high $\Mdot$ state during the HST observation, and we take the rather
high WD luminosity as an upper limit to the accretion luminosity.

Note that we are making the unconventional assumption that the white
dwarf luminosity is the accretion luminosity in the low state.  The
consequences of other assumptions are discussed in \S 4.
ADAF models predict $L_{wd} = L_{acc}$.  Conventional boundary
layer models (Pringle \& Savonije 1979; Tylenda 1981) predict only
$L_{wd} \sim 0.5 L_x$ from the heated surface of the white
dwarf, although some models of optically thick boundary layers advect
a significant fraction of the luminosity into the star (Regev 1983).
The soft X-ray excess of some AM Her stars is a clear, if poorly
understood, example of advection of accretion energy into the white
dwarfs in magnetic CVs.

The WDs in dwarf novae are heated during outbursts, and cool during
the quiescent phase.  Cooling times are fairly short, with values of
30 days and 6 days estimated for U Gem and RX And, respectively (Long
et al 1993; Sion 1999).  Compressional heating throughout the WD (Sion
1995) reflects the long-term average accretion rate.  We have no
reliable way to remove the effects of heating during outburst or
compressional heating, though we have tried to select WD temperatures
far after outburst when possible.  Therefore, it must be borne in mind
that some fraction of $L_{wd}$ may arise from causes other than
accretion during the quiescent phase. However, as noted in \S 1,
factors of $\sim2$--$3$ have little effect on our quantitative
conclusions.

\subsection{Neutron Stars}

Table 1 shows the 0.5--10 keV X-ray luminosities of all NS SXTs for
which data are available.  The luminosity of SAX J1808.4-3658 is from
Stella et al. (2000), and those of the other four sources are from
Narayan et al. (1997) and Menou et al. (1999b).  The latter authors
give the luminosity of one other NS SXT, H1608-52, which has
$\log[(L_x/{\rm erg~s^{-1}})]=33.3$.  However, 
this source does not have a reliably
measured orbital period; the period could be either 5 hours (Chen et
al. 1998) or 98.4 hours (Ritter \& Kolb 1998).  Since it is important
for our method to select systems with comparable orbital periods (so
that the outer boundary conditions for the accretion flows are
similar), we do not include this system.

The values tabulated in Table 1 may be regarded as estimates of the
total (bolometric) luminosities of the systems since a substantial
fraction of the emission is expected to come out in the 0.5--10 keV
band.  In the case of Cen X-4 in quiescence, McClintock \& Remillard
(2000) have measured the optical-UV spectrum and shown that the
luminosity in these bands is comparable to the X-ray luminosity.
Assuming that this source is typical, we expect the 0.5--10 keV
luminosity to underestimate the bolometric luminosity by a factor of 2
or so.

A more difficult issue is deciding what fraction of the X-ray
luminosity is derived from accretion.  Recently, Brown, Bildsten \&
Rutledge (1998) showed that crustal heating of NSs in SXTs during
outburst could lead to quiescent thermal emission from the cooling NSs
at a level comparable to the observed luminosities of quiescent NS
SXTs.  Could the entire observed $L_x$ be due to this cooling
radiation?

At least two of the five NS SXTs listed here have shown substantial
variability: Cen X-4 (Campana et al. 1997) and SAX J1808.4-3658
(Dotani, Asai \& Wijnands 2000).  Such variability is not expected for
a cooling NS.  Also, the X-ray spectra of two quiescent systems show
significant power-law tails: Cen X-4 (Asai et al. 1996) and Aql X-1
(Campana et al. 1998).  Again, it is unlikely that the power-law
emission arises from thermal cooling.  Barring the somewhat extreme
possibility that quiescent NS SXTs behave similarly to radio pulsars
and have significant magnetospheric activity (as proposed by Campana
\& Stella 2000), we feel that there is a strong case for assuming that
at least the power-law emission (which has roughly half the power) is
due to accretion.  Thus, the accretion luminosities of the NS SXTs
considered are probably between $\sim 50\%$ and 100\% of the
luminosities listed in Table 1 (50\% if only the power-law spectral
component is from accretion, and 100\% if the thermal component is
also from accretion).  In Figure 1 and the analysis presented in \S3, we
assume 100\%.  (Equivalently, we assume 50\% and apply a bolometric
correction of a factor of 2.)

\section{Inferred Mass Accretion Rates and Estimate of $p$}

Using the accretion luminosities in Table 1, we may now estimate the
mass accretion rates from equation~(\ref{eq:1}). We assume that all
NSs have the same mass, $M_\star\approx 1.4 M_\odot$ (Thorsett \&
Chakrabarti 1999), and radius, $R_\star \approx 10^6~{\rm cm}$
(Shapiro \& Teukolsky 1983).  For the WDs, we use the masses and radii
given in Table 1.

Figure 1 shows two clusters of points corresponding to WD and NS
systems in quiescence. The three lines show power-law scalings with
different slopes, $p=0.4, 0.9$ and $1.4$, all originating from the
logarithmic average of $\Mdot_\star$ for NS systems. Since the WD
points are roughly bounded by these lines we infer that $\Mdot_\star
\propto R_\star^{0.9\pm 0.5}$ for accretion flows in quiescence around
compact stars. The central value of $p$ is close to unity, in
agreement with the expected behaviour of convection-dominated
accretion flows (CDAF).

\section{Conclusions}

The accretion luminosities of NSs and WDs in quiescence with similar
binary companions are comparable (see Table 1). This surprising result
implies that the accretion rate at the surface of a WD, namely
$\Mdot_\star(10^9~{\rm cm})$, is larger by three orders of magnitude
than that on the surface of a NS, $\Mdot_\star(10^6~{\rm cm})$.  In
both cases the radius of the hard surface where the accretion flow
terminates is much smaller than the outer boundary of the accretion
flow, and hence the flow should exhibit a similar behaviour on scales
larger than its inner edge.  Figure 1 implies that the mass accretion
rate varies with radius roughly as $\Mdot_\star \propto
R_\star^{0.9\pm 0.5}$. This behaviour is remarkably similar to that
predicted for CDAFs (Narayan et al. 2000; Quataert \& Gruzinov 2000;
Igumenshchev et al. 2000).  The data are marginally consistent with an
ADIOS model (Blandford \& Begelman 1999), for which Igumenshchev \&
Abramowicz (1999, 2000) found $p\sim0.3-0.5$.  The data do not support
the assumption of a constant $\Mdot_\star$, as in a pure ADAF model
(e.g. Narayan et al. 1996), or a very steep scaling ($p\sim2.7$) as in
a pure quiescent thin disk (e.g. Menou et al. 1999a).

Unfortunately, the quantitative results are subject to large uncertainties
due to the small number statistics of the relevant systems.  They could
also be affected by a number of systematic effects.  One potential
systematic effect is probably not important, namely we do not believe that
there is a serious bias in the sample of systems we have used.  Our sample
of CVs is an unbiased sample of binaries with orbital periods in the range
of interest.  Also, virtually all known NS SXTs have been detected in
X-rays in quiescence; therefore, the sample of NS SXTs in Table 1 is likely
to be complete and unbiased (except for the one system we did not include
because its period is not known, see \S2.2).

One potentially important systematic effect is related to the fact that the
luminosities in Table 1 may not represent just the accretion luminosities
in quiescence, but may also include a cooling component which originates
from the heat deposited in the accretor during an earlier high $\Mdot$
(outburst) state.  If some of the quiescent emission in CVs is due to the
cooling of the WD (Sion 1999), then $p$ would be smaller than the value we
obtained; in principle, if the accretion luminosity in quiescent CVs is
very much smaller than the observed luminosity, $p$ could be as small 
as zero,
as required for a pure ADAF model.  The luminosity of a cooling WD should
decay with time in a predictable fashion, so this scenario could be tested
with careful observations of the CVs listed in Table 1.  On the other hand,
if the quiescent luminosities of NS SXTs are primarily due to NS cooling
(as proposed by Brown et al. 1998), then $p$ would be larger than our
estimated value, perhaps as large as the 2.7 predicted for a pure thin
disk.  For the reasons given in \S2.2, we consider this unlikely.

A systematic effect with an opposite sign involves the possibility
that we have underestimated the bolometric luminosities of the systems
under consideration by ignoring the emission at unobserved photon
frequencies.  This is probably not a serious problem, as CVs appear to
radiate primarily in optical, UV and EUV, where there are adequate
observations, and NS SXTs emit primarily in the
X-ray band (see \S2.1 and \S2.2).

Finally, it is possible that magnetic fields of NSs or WDs reduce the
mass accretion rate onto the star, thereby giving an artificially low
estimate of $\Mdot_\star$.  This is more likely in NS SXTs than in CVs
(Menou et al. 1999b).  If the effect is important, then $p$ would be
smaller than our estimate.

A key aspect of our analysis is the relation between the mass
accretion rate and the luminosity given in equation (2).  This formula
is valid so long as the accreting star has a hard surface.  If the
accretor is a black hole, and if accretion occurs via a radiatively
inefficient flow, such as an ADAF or an ADIOS or a CDAF (but not a
thin disk), then the accretion luminosity is expected to be
significantly less than that given in equation (2).  This is because
the energy advected with the accretion flow can disappear through the
event horizon (Narayan \& Yi 1995b; Narayan et al. 1997).  The
quiescent luminosities of BH SXTs do appear to be significantly lower
than the luminosities of NS SXTs.  This is consistent with the
presence of an event horizon in black hole SXTs (Narayan et al. 1997;
Menou et al. 1999b).

\acknowledgements

This work was supported in part by NASA grants NAG 5-7039 and NAG
5-7768, NSF grants PHY-9507695 and AST-9900077, and by a grant from
the Israel-US BSF for AL.

\newpage

\begin{table}[h]
\begin{center}
\centerline{Table 1: Accretion Luminosities in Quiescence (in ${\rm
erg~s^{-1}}$)} \scriptsize
\vskip 20pt
\begin{tabular}{|l c c c c c c c c c|}
\hline
\multicolumn{10}{|c|}{White Dwarf (CVs)} \\
\hline
Object & $P$(hr)& $M_{wd}(M_\odot)$ & log ($T_{wd}$/K) & V & d(pc)& log($R_{wd}$/cm)& log $L_x$& log $L_v$  & log ($L_{wd}$)\\

\hline
RX And & 5.02 & 1.14$\pm$0.33  & 35,000 & 12.6  &   135  & 8.63     &  30.0  &  31.9     &    32.3       \\
SS Cyg    &   6.60 &   1.19$\pm$0.02  &   37,000   & 11.4  &   166 & 8.62 &  32.6    &  32.6     &    33.0      \\
U Gem     &   4.25 &   1.02$\pm$0.04  &   30,000*  & 14.0  &    96  & 8.75&  31.2    &  31.5     & 32.6       \\
IP Peg    &   3.80 &   1.15$\pm$0.10  & $<15,000$  & 17.8  &   124 &    8.63   &$<29.5$   &  30.4        & 30.8       \\
TZ Per    &   6.31 &    (1.0)        &   18,000   & 15.6  &($>380$) &   (8.76) &($>30.4$) &  31.6         & 31.4         \\
DW UMa    &   3.28 &   0.48$\pm$0.06  &   46,000   & 17.6  &   830 &    9.11 &          &  31.4          & 33.7        \\
\hline
\end{tabular}
\end{center}
\begin{center}
\begin{tabular}{|l c c|}
\hline
\multicolumn{3}{|c|}{Neutron Star (SXTs)}  \\
\hline
\hline
Object  &  P (hr)  & Log ($L_x$[0.5-10 keV])  \\
\hline
SAX J1808.4-3658  &  2 & 32.4 \\
EXO 0748-676      &  3.8 & 34.1\\
4U2129+47         &  5.2 & 32.8\\
1456-32 (Cen X-4) &  15.1 & 32.4\\
1908+005 (Aql X-1) &  19 & 32.6\\
\hline
\end{tabular}
\end{center}
\end{table}
\normalsize
\newpage
\begin{figure}[p]
\includegraphics{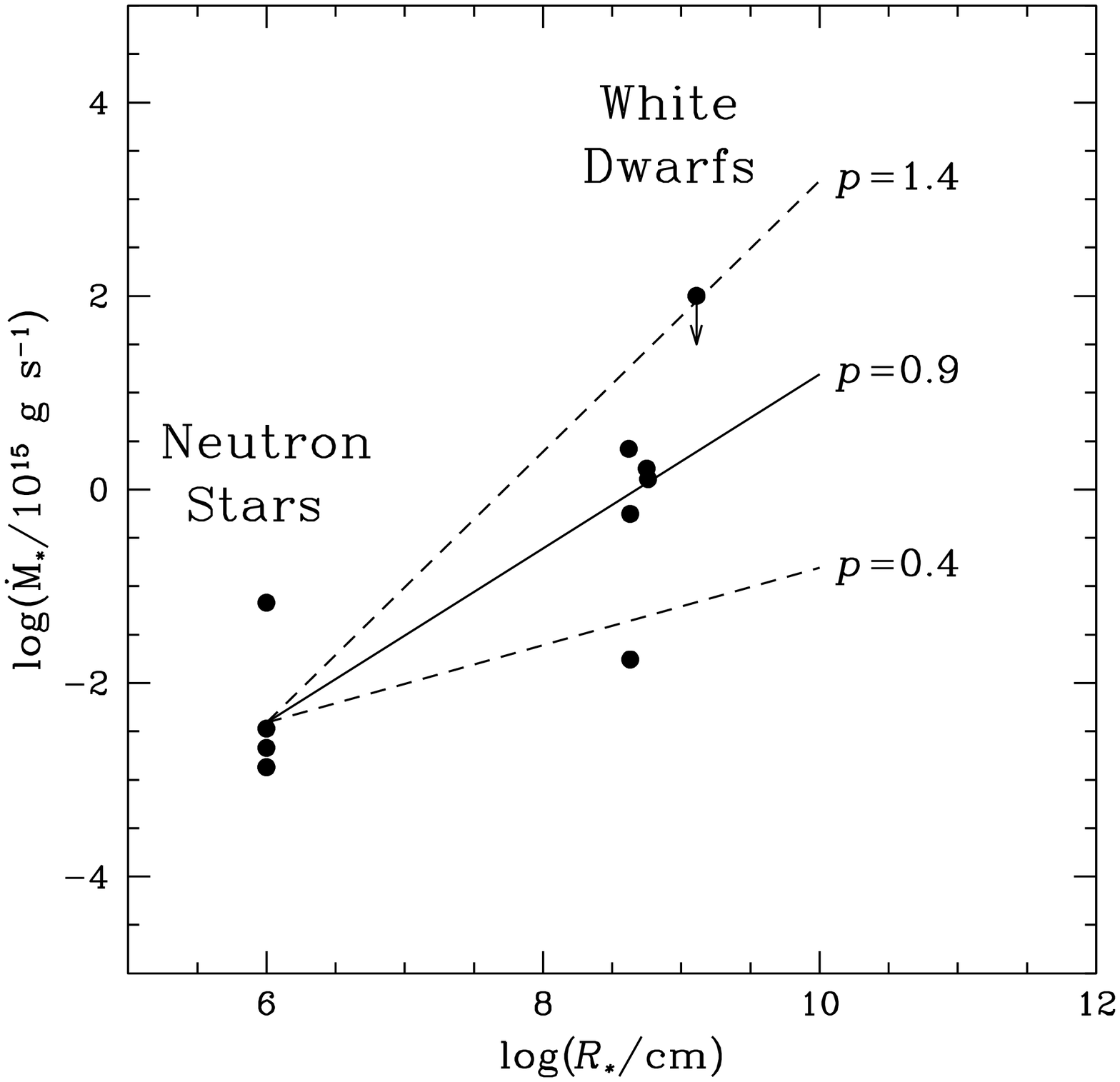}
\vspace{6.2in}
\caption{Mass accretion rate vs radius, $\Mdot_\star(R_\star)$, 
for white dwarfs (WD)
and neutron stars (NS) which accrete mass from binary companions with
similar orbital periods. The points were inferred from the data in Table 1
using Eq. (2). The solid line originates at the (logarithmic) average of
$\Mdot_\star$ for the NS points and goes through the 
average $\Mdot_\star$ for the WD
points. It corresponds to a power-law scaling with an index $p=0.9$
[Eq. (1)], while the dashed lines correspond to $p=1.4$ (upper line) and
$p=0.4$ (lower line).}
\label{fig-1}
\end{figure}


\begin{references}

\reference{} Abramowicz, M. A., Chen, X., Kato, S., Lasota, J.-P., \&
Regev, O.  1995, ApJ, 438, L37

\reference{} Asai, K., Dotani, T., Mitsuda, K, Hoshi, R., Vaughan, B.,
Tanaka, Y., \& Inoue, H. 1996, PASJ, 48, 257

\reference{} Blandford, R. D., \& Begelman, M. C. 1999, MNRAS, 303, L1

\reference{} Bondi, H. 1952, MNRAS, 112, 195

\reference{} Campana, S., \& Stella, L. 2000, ApJ, 541, 849

\reference{} Campana, S., Mereghetti, S., Stella, L., \& Colpi,
M. 1997, A\&A, 324, 941

\reference{} Campana, S. et al., 1998, ApJ, 499, L65

\reference{} Chen, X., Abramowicz, M. A., Lasota, J.-P., Narayan, R., \&
Yi, I. 1995, ApJ, 443, L61

\reference{} Dotani, T., Asai, K., \& Wijnands, R. 2000, ApJL, in
press; astro-ph/0009295

\reference{} Froning, C.S., Robinson, E.L., Welsh, W.F., \& Wood,
J.H. 1999, ApJ, 523, 399.

\reference{} Gruzinov, A. 1998, ApJ, submitted; astro-ph/9809265

\reference{} Hamada, T., \& Salpeter, E.E. 1961, ApJ, 134, 683.

\reference{} Hameury, J-M., Menou, K., Dubus, G., Lasota, J-P., \& Hure,
J-M.  1998, MNRAS, 298, 1048

\reference{} Harrison, T.E., McNamara, B.J., Szkody, P., McArthur, B.E.,
Benedict, G.F., Klemola, A.R., \& Gilliland, R.L. 1999, ApJL, 515, L93

\reference{} Ichimaru, S. 1977, ApJ, 214, 840

\reference{} Igumenshchev, I.V., \& Abramowicz, M.A. 1999, MNRAS, 303,
309

\reference{} Igumenshchev, I.V., \& Abramowicz, M.A. 2000, ApJ, in
press; astro-ph/0003397

\reference{} Igumenshchev, I. V., Abramowicz, M. A., \& Narayan, R.  2000,
ApJ, 537, L27
 
\reference{} Knigge, C., Long, K.S., Hoard, D.W., Szkody, P., \& Dhillon,
V.S. 2000, preprint

\reference{} Lasota, J.-P., Narayan, R., \& Yi, I. 1996, A\&A, 314,
813

\reference{} Long, K.S. 2000, New Astronomy Reviews, 44, 125

\reference{} Long, K.S., Blair, W.B., \& Raymond, J.C. 1993, ApJ
Lett., 454, L39.

\reference{} Ludwig, K. Meyer-Hofmeister, E. \& Ritter, H. 1994, A\&A,
290, 473

\reference{} McClintock, J.E., \& Remillard, R.A. 2000, ApJ, 531, 956
 
\reference{} Medvedev, M., \& Narayan, R. 2000, ApJ, submitted;
astro-ph/0007064

\reference{} Menou, K., Narayan, R., \& Lasota, J.-P. 1999a, ApJ, 513,
811

\reference{} Menou, K., Esin, A. A., Narayan, R., Garcia, M. R., Lasota,
J.-P., \& McClintock, J. A. 1999b, ApJ, 520, 276

\reference{} Meyer, F., \& Meyer-Hofmeister, E. 1994, A\&A, 288, 175

\reference{} Narayan, R., Garcia, M. R., \& McClintock, J. E.  1997, ApJ,
478, L79

\reference{} Narayan, R., Igumenshchev, I. V., \& Abramowicz, M. A. 
2000, ApJ, 539, 798

\reference{} Narayan, R., McClintock, J. E, \& Yi, I. 1996, ApJ, 457,
82

\reference{} Narayan, R., \& Yi, I. 1994, ApJ, 428, L13

\reference{} ---------------------------------. 1995a, ApJ, 444, 231

\reference{} ---------------------------------. 1995b, ApJ, 452, 710

\reference{} Novikov, I., \& Thorne, K. S. 1973, in Black Holes,
ed. C. de Witt, \& B. de Witt (New York: Gordon \& Breach), 422

\reference{} Pringle, J.E., \& Savonije, G.J. 1979, MNRAS, 187, 777.
 
\reference{} Quataert, E., \& Gruzinov, A. 2000, ApJ, 539, 809

\reference{} Regev, O. 1983, A \& A, 126, 146.

\reference{} Ringwald, F.A. 1995, MNRAS, 274, 127

\reference{} Ritter, H., \& Kolb, U. 1998, A\&AS 129, 83

\reference{} Sion, E.M. 1995, ApJ, 438, 876.
 
\reference{} ----------------------. 1999, PASP, 111, 532.

\reference{} Shakura, N. I., \& Sunyaev, R. A. 1973, A \& A, 24, 337
 
\reference{} Shapiro, S. L., \& Teukolsky, S. A. 1983, Black Holes, White
Dwarfs, \& Neutron Stars, Wiley: New York, Ch. 9

\reference{} Stella, L., Campana, S., Mereghetti, S., Ricci, D., \&
Israel, G.L. 2000, ApJ, 537, L115

\reference{} Stone, J. M., Pringle, J. E., \& Begelman, M. C. 1999,
MNRAS, 310, 1002
 
\reference{} Szkody, P., Long, K.S., Sion, E.M., \& Raymond, J.C. 1996,
ApJ, 469, 834.

\reference{} Thorsett, S. E., \& Chakrabarty, D. 1999, ApJ, 512, 288

\reference{} Tylenda, R. 1981, Acta Astr., 31, 267.
 
\reference{} Verbunt, F., Bunk, W.H., Ritter, H., \& Pfeffermann, E. 1997,
A\&A, 327, 602
 
\reference{} Yoshida, K., Inoue, H., \& Osaki, Y. 1992, PASJ, 44, 537.
 

\end{references}
\end{document}